# Optical Cooper Pair Breaking Spectroscopy of Cuprate Superconductors

Y. G. Zhao*, Eric Li, Tom Wu, S. B. Ogale, R. P. Sharma, and T. Venkatesan
Center for Superconductivity Research, Department of Physics, University of Maryland, College Park, MD 20742
J. J. Li, W. L. Cao, C. H. Lee
Department of Electrical Engineering, University of Maryland
College Park, MD 20742
H. Sato, and M. Naito
NTT Basic Research Laboratories, 3-1 Morinosato Wakamiya, Atsugi-shi
Kanagawa 243-0198, Japan

The photon energy dependence of optical Cooper pair breaking rate (CPBR) is studied for compressibly strained $La_{1.85}Sr_{0.15}CuO_4$ (LSCO) films and $YBa_2Cu_{2.92}Zn_{0.08}O_{7-\delta}$ (YBCZO) thin films, and compared to that in $YBa_2Cu_3O_{7-\delta}$ (YBCO). Unlike in the case of YBCO, the CPBR for LSCO does not show obvious photon energy dependence. In YBCZO the CPBR shows strong photon energy dependence similar to YBCO, but with a red shift in the peak position. Analysis of these results strongly favours a physical picture based on electronic phase separation in high $T_c$ superconductivity.

The microscopic mechanism of high temperature superconductivity (HTS) remains a mystery till todate, although significant understanding has been developed over the years in elucidating the key underlying factors. Various models have been proposed for the understanding of HTS.[1] It is commonly believed that strong correlation between electrons play a very important role in this system, however, the manner in which such correlation unfold as collective behavior is still not understood. Recently, experimental evidence is accumulating in favor of the occurrence of electronic phase separation (EPS) in such strongly correlated systems; the so-called stripe phase picture being one manifestation of such a scenario.[2] The EPS picture implies an inhomogeneity of both the charges and spins in HTS. It is not yet clear whether the EPS or stripes are central to the phenomenon of high $T_C$ superconductivity.

Based on thermal difference reflectance (TDR) spectroscopy work on cuprates, Little et al.[3] concluded that phonons and a high energy electronic excitation (ranged over 1.6 eV to 2.3 eV) are jointly important for pairing in HTS. Stevens et al.[4] performed pump-probe measurement on YBCO employing excitation by 3 eV photons, with the probe beam detecting the excited state and its relaxation. Their results also showed an absorption peak around 1.5 eV, broadly consistent with the TDR measurement, however their interpretation of the origin of the peak differs from that of Little et al.[3] In our previous work[5] on electrically characterized optical pair breaking (which differs distinctly from the all-optical measurements by other researchers), we observed a fairly sharp peak in the Cooper pair breaking rate around 1.5 eV for YBCO, confirming a resonance. Existence of such a sharp feature is indeed surprising if one were to think of the superconductor as a uniform conductor. Noting that all these works reflect importance of states separated in energy by about ~1.5 eV we decided to probe the case further by examining other cuprate systems, namely LSCO and Zn-doped YBCO. We find that over the energy range in which YBCO shows the resonance, LSCO does not show any obvious photon energy dependence. On the other hand, $YBa_2Cu_{2.92}Zn_{0.08}O_{7-\delta}$ (YBCZO) does show a strong photon energy dependence, but with a red shift of the peak feature. We argue that it is difficult to reconcile all these data without invoking the electronic phase separation picture for HTS cuprates.

The $YBa_2Cu_{2.92}Zn_{0.08}O_{7-\delta}$ thin films were prepared by pulsed laser deposition (PLD) on (100) $LaAlO_3$ substrates. The thickness of the films was about 100 nm, with $T_C$ ~ 58 K. $La_{1.85}Sr_{0.15}CuO_4$ thin films were prepared by reactive co-evaporation (electron beam evaporation) on (001) $LaSrAlO_4$ substrates following Sato et al.[6] The thickness of the films was about 100 nm with $T_c$ between 40 and 43 K. The films were patterned by standard photolithographic technique to obtain the coplanar waveguide devices. The patterning process decreased the $T_c$ of LSCO films to 34 K. The sketch of the experimental setup and the device (essentially an optically controlled opening switch) can be found in our previous paper.[7] The device was mounted on a cold finger located in a vacuum cryogenic chamber and biased with a dc current. The device was illuminated with 100 femtosecond pulses from a Ti:Sapphire laser system including an oscillator and a regenerative amplifier with an ability to deliver 5 μJ/pulse at a repetition rate of 9 kHz. The high peak power and suitable repetition rate allow efficient fast switching without thermal heating problems, as discussed earlier.[5,7] The wavelength of the laser was tunable within the range of 760-860 nm (1.63-1.44 eV). When the ultrashort laser pulse illuminated the bridge, transient switch current waveforms were produced instantaneously, resulting in a fast drop of current flowing through the device. These waveforms were monitored by a fast sampling oscilloscope with a temporal resolution of 20 ps. In the experiment, great care was taken to keep the laser power constant and the beam focused on the superconducting bridge.

Fig. 1 shows the typical waveform of the fast optical response for LSCO films. The rise time and fall times of the signal are around 40 ps. This waveform is similar to that for YBCO.[5] It has been established that this signal is related to the Cooper pair breaking,[8,9] which changes the kinetic inductance of the superconducting waveguide. The amplitude of this signal can be expressed as $V = IR (\Delta L_{kin}/\Delta t) / (2\Delta L_{kin}/\Delta t+4R)$,[5,7] where I is the bias current, R is 50Ω, $\Delta t$ is the pair breaking time, $L_{kin}$ is the kinetic inductance of the superconducting bridge, and $\Delta L_{kin}/\Delta t = (m^*l/(e^2wdn_s^2))(\Delta n_s/\Delta t)$. In this formula, $m^*$, $n_s$ and e are the effective mass, the density, and the charge of superconducting carriers, respectively. Parameters d, l, w are the thickness, length, and width of the bridge, respectively. Thus, from the amplitude V measured using a fast oscilloscope, we can obtain $\Delta L_{kin}/\Delta t$. Since $\Delta L_{kin}/\Delta t$ is proportional to the pair breaking rate $\Delta n_s/\Delta t$, the temperature and photon energy dependence of the pair breaking rate can be studied. In such an argument we assume that m* is fixed. However, in the stripe phase picture, this aspect may have to be reexamined.

Fig. 2 shows the temperature dependence of $\Delta L_{kin}/\Delta t$ for LSCO films. The behavior is similar to that in YBCO, and can be explained qualitatively by using the two fluid model.[5] Note that $\Delta L_{kin}/\Delta t$, as given by $(ml/(e^2 w d n_s^2))(\Delta n_s/\Delta t)$, is essentially proportional to $\Delta n_s/\Delta t/(n_s^2)$, since m, l, e, w and d are constants. If $\Delta n_s/\Delta t$ is temperature independent or weakly temperature dependent, the temperature dependence of the amplitude of $\Delta L_{kin}/\Delta t$ will be determined by $1/n_s^2$. Since $n_s$ increases with decreasing temperature, $\Delta L_{kin}/\Delta t$ is expected to be reduced rapidly as the temperature decreases. The inset of Fig. 2 shows the square root of $1/\Delta L_{kin}/\Delta t$, which is proportional to $n_s$, if $\Delta n_s/\Delta t$ is temperature independent or weakly temperature dependent. The temperature dependence of $n_s$ shown here is different from the $n_s(T)$ curve reported by Hardy et al.,[10] which shows a linear temperature dependence of $n_s$ at low temperature, consistent with the d-wave pairing mechanism. This discrepancy implies that $\Delta n_s/\Delta t$ has some temperature dependence. Indeed, it has been shown that the charge transfer (O2p to Cu 3d) gap, which is related to the photon absorption, increases with temperature[11] and the life time of the quasiparticles produced by the Cooper pair breaking process is also expected to change with temperature.[12] If we use the $n_s(T)$ data obtained from other experiments,[10] it is possible to estimate the T dependence of $\Delta n_s/\Delta t$.

Fig. 3(a) gives the photon energy dependence of $\Delta L_{kin}/\Delta t$ for LSCO thin films. Unlike YBCO, it does not show any noticeable photon energy dependence. The YBCO data are redrawn in fig. 3(c) for comparison. This indicates that the resonance of Cooper pair breaking observed in YBCO [5] is intrinsic. Fig. 3(b) gives the photon energy dependence of $\Delta L_{kin}/\Delta t$ for $YBa_2Cu_{2.92}Zn_{0.08}O_{7-\delta}$. It shows dramatic photon energy dependence as was seen in YBCO. It is clear that the resonance peak shifts to lower energies as compared to that of the resonance peak in YBCO shown in Fig. 3(c).[5]

Now we turn to the analysis of our results. In reference [3], the high energy electronic excitation (~1.5-1.7 eV), which is suggested to be related to the pairing in HTS, is attributed to the energy of the $d^9$-$d^{10}L$ charge transfer excitation associated with the $CuO_2$ network which is common to HTS systems. This charge transfer excitation has also been observed in superconducting YBCO by Electron Energy Loss Spectra.[13] However, the feature we observed in the CPBR spectrum of YBCO near 1.5 eV is considerably narrower (100 meV) than that (500 meV) of the peak in reference [3]. We argue that the observation of such a sharp resonance is hard to understand for any homogeneous conducting state. On the other hand, presence of insulating regions in the superconducting state, as is envisaged in the EPS or Stripe phase scenario, can lead to narrow absorption features provided that the absorption-induced perturbation of the insulating (antiferromagnetic) state directly couples with the paired hole system and breaks pairs. Interestingly, the insulating $YBa_2Cu_3O_6$ compound has the charge transfer excitation peak (from the O2p to Cu3d upper Hubbard band) around 1.7 eV. [14-16] In the small phase separation length-scale anticipated in the EPS or stripe scenario there could be renormalization of the energy of this peak causing its shift to lower energy. To what extent is the charge transfer peak for the insulating domains/stripes in superconducting $YB_2Cu_3O_{7-\delta}$ different from that of the insulating bulk $YB_2Cu_3O_6$, is still an open question, which needs theoretical inputs. In the related context it is useful to point to a recent observation that the screening of phonon modes in high $T_c$ superconductors is poor or totally absent, and the majority of the phonon modes have oscillator strengths similar to those found in the insulating materials.[17] Therefore, it is reasonable to expect that photons are absorbed mainly by the insulating domains/stripes rather than the metallic ones, and that the insulating domains/stripes dominate the optical properties of high $T_c$ superconductors.

In our experiment we selectively and electrically probe the broken Cooper pairs in an ultrafast measurement. The speed and the concept of our measurement are key to the results we obtain. Kataev et al studied the temperature dependence of the spin fluctuation frequency for $La_{2-x}Sr_xCuO_4$ samples by ESR of Gd spin probes.[18] The spin fluctuation frequency shows strong temperature dependence and changes from $3\times10^{13}$ Hz at 250 K to about $10^{10}$ Hz at 5 K. Therefore, the time scale for the spin fluctuation is $10^{-13}$ s at high temperature and $10^{-10}$ s at low temperature. In our experiment, the width of laser pulse is only 100 fs ($10^{-13}$ s), which is very fast in comparison with the spin fluctuation time scale. If the latter is considered to represent stripe fluctuations, our measurement would essentially reflect the snapshot picture of charge/spin domains or stripes at a certain time. In contrast, the measurement in ref. [3] represents a time average. This time scale difference could be a factor responsible for the different widths obtained in our experiment and the TDR experiment.

As discussed above, a reasonable explanation for the CPBR in the case of YBCO is the charge transfer excitation in spatially confined domains/stripes of antiferromagnetic (AF) insulating regions in YBCO. It is possible that a similar excitation in the AF insulating regions of LSCO is out of the photon energy range employed in this work. For example, the charge transfer energy for $La_2CuO_4$ is about 2 eV,[14] which is higher than the charge transfer energy of 1.7 eV for $YBa_2Cu_3O_6$.[15, 16] Hence, even after renormalization and shift, it may not fall in the range of the measurement. An alternate proposal for the absence of the CPBR in the case of the 214 film could be that both static and dynamic spin/charge stripes are absent in the compressibly strained LSCO thin films.[19] At this stage the existence of the dynamic stripes is still an open issue. Further work is clearly needed to extend the photon energy to both lower and higher energy sides, especially close to the 2 eV CT gap of $La_2CuO_4$, to verify whether CPBR exists in LSCO and hence a similar conclusion as for YBCO can be drawn for the 214 case as well.

For Zn doped $YB_2Cu_3O_{7-\delta}$, even though Zn is expected to be in a non-magnetic $3d^{10}$ state, its destruction of superconductivity is even stronger than magnetic ion, such as Ni.[20,21] It has been found that Zn doping induces magnetic moment on Cu sites around Zn,[22-24] and that this moment couples strongly with the conduction band at low temperature.[25] Charge localization was reported in Zn doped $YB_2Cu_3O_{7-\delta}$ and has been explained by the destruction of the local AF correlation among Cu spins by Zn.[26] However, recent NMR result suggests that the AF correlations are enhanced rather than destroyed around Zn.[27] Therefore, other scenario is needed to explain the localization effect. It is also suggested that Zn impurities are surrounded by extended regions whose magnetic properties are strongly modified already far above $T_c$, and wherein superconductivity never develops.[28] Superconductivity is then confined to regions far from the Zn impurities. For Zn doped $Bi_2Sr_2CaCu_2O_{8+\delta}$, STM study also shows that superconductivity is strongly suppressed within 1.5 nm of the scattering sites.[29] In the stripe phase model, superconductivity is related to the fluctuation of the stripes.[30] It has been suggested that the pinning of the dynamically fluctuating stripes results in the suppression of superconductivity.[31] In our experiment, Zn doping should not affect the results very much in terms of the time scale since the stripes are static to the probing light pulse even for the undoped YBCO because of our ultrafast technique. However, Zn doping may affect the charge transfer gap because of the suggested modification to the bands, which leads to the shift of the CPBR resonance peak to lower energy.

In summary, we have studied the photon energy dependence of the Cooper pair breaking rate (CPBR) for $La_{1.85}Sr_{0.15}CuO_4$ and $YBa_2Cu_{2.92}Zn_{0.08}O_{7-\delta}$ thin films, and compared them with that in YBCO. The strong photon energy dependence of CPBR in YBCO and YBCZO (with a redshift), and its absence in LSCO strongly favour the electronic phase separation (or stripe) picture for cuprates; the absorption responsible for the measured pair breaking being the charge transfer excitation in the insulating antiferromagnetic domains confined between charge lines.

We would like to acknowledge the support from ONR Grant No. ONR-N000149611026.

* Present address: Department of Physics, Tsinghua University, Beijing 100084, P. R. China.

Figure Captions

Fig.1. The waveform of the fast optical transient signal related to the Cooper pair breaking.

Fig.2. Temperature dependence of $\Delta L_{kin}/\Delta t$ for $La_{1.85}Sr_{0.15}CuO_4$ thin film. The inset shows the square root of $1/\Delta L_{kin}/\Delta t$, which is proportional to $n_s$ if we assume $\Delta n_s/\Delta t$ is temperature independent or weakly temperature dependent.

Fig.3. Photon energy dependence of $\Delta L_{kin}/\Delta t$ for (a) $La_{1.85}Sr_{0.15}CuO_4$ (b) $YBa_2Cu_{2.92}Zn_{0.08}O_{7-\delta}$ and (c) $YBa_2Cu_3O_{7-\delta}$ thin films (Y. G. Zhao et al., J. Super. 12, 675 (1999)).

.

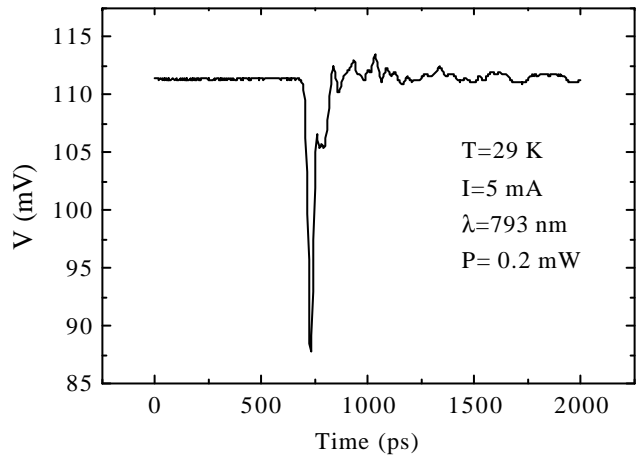

Fig. 1

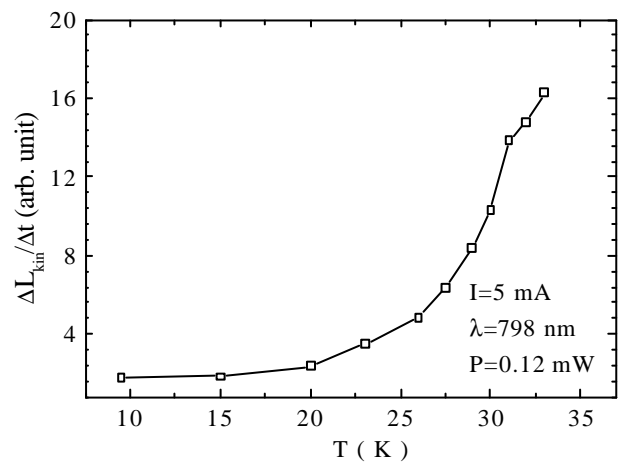

Fig.2

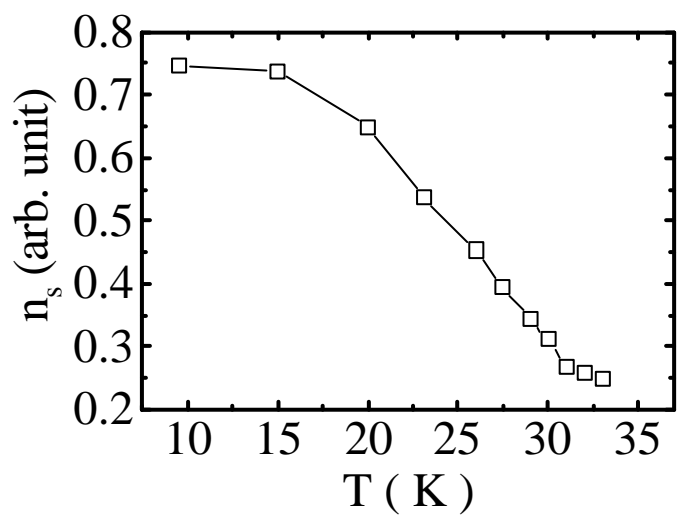

Inset of fig.2

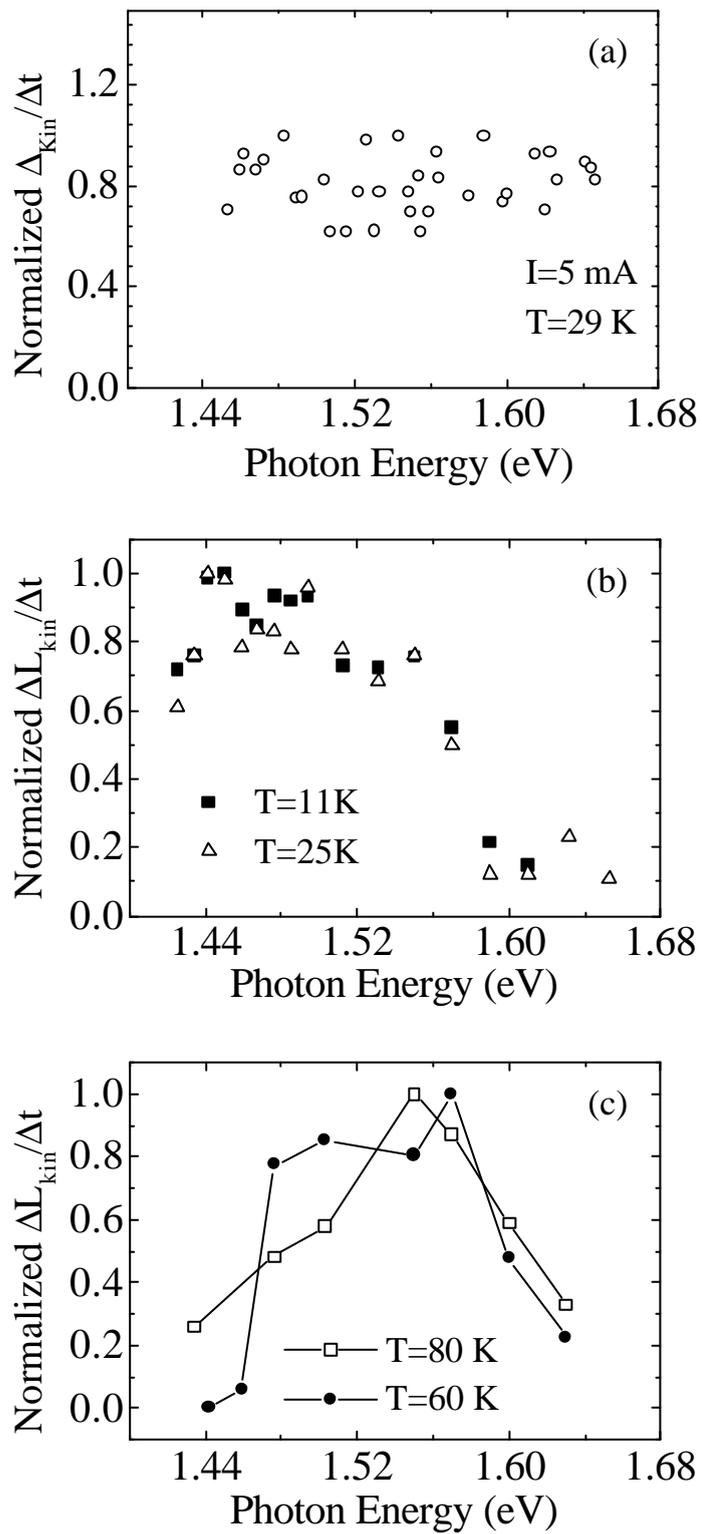

Fig. 3